\documentclass[12pt]{article}
\usepackage{latexsym}
\usepackage{graphics}
\usepackage{epsfig}
\usepackage{epstopdf}
\usepackage{amsmath}
\usepackage{cite}
\usepackage{hyperref}
\usepackage{amssymb}

\newcommand{\be}{\begin{equation}}
\newcommand{\ee}{\end{equation}}
\newcommand{\beq}{\begin{eqnarray}}
\newcommand{\eeq}{\end{eqnarray}}
\newcommand{\beqn}{\begin{eqnarray*}}
\newcommand{\eeqn}{\end{eqnarray*}}
\newcommand{\ba}{\hspace*{-5pt}\begin{array}}
\newcommand{\ea}{\end{array}}

\newcommand{\bit}{\begin{itemize}}
\newcommand{\eit}{\end{itemize}}
\newcommand{\ben}{\begin{enumerate}}
\newcommand{\een}{\end{enumerate}}
\newcommand{\prf}{{\bf Proof: }}
\newcommand{\eprf}{\rule{7pt}{7pt}}

\newtheorem{theo}{Theorem}

\newtheorem{lem}{Lemma}
\newtheorem{prop}{Proposition}
\newtheorem{cor}{Corollary}
\newtheorem{rem}{Remark}

\begin{document}

\begin{center}
\textbf{\Large Existence of the solitary wave solutions supported by the hyperbolic  modification of the FitzHugh-Nagumo system} \vspace{0.5 cm}

 Aleksandra Gawlik$^{a,}$\footnote{e-mail:
\url{aleksandramalgorzatagawlik@gmail.com}}, Vsevolod  Vladimirov$^{a,}$\footnote{e-mail:
\url{vladimir@mat.agh.edu.pl}}, Sergii  Skurativskyi$^{b,}$\footnote{e-mail: \url{
skurserg@gmail.com}} \vspace{0.5 cm}

{\small \it
$^a$ Faculty of Applied Mathematics, \\
AGH University of Science and Technology, \\
Mickiewicz Avenue 30,  30-059 Krak\'{o}w, Poland,

$^b$ Division of Geodynamics of Explosion, \\
 Subbotin Institute of Geophysics, NAS of Ukraine,  \\
Acad. Palladina Avenue 32,  03142 Kyiv, Ukraine}

%Subbotin institute of geophysics, Nat. Acad. of Sci. of Ukraine

   %  Bohdan Khmelnytskyi str. 63-G, Kyiv, Ukraine
%
\end{center}

\begin{quote} \textbf{Abstract. }{\small 
 We study a system of nonlinear differential equations simulating transport phenomena in active media. The model we are interested in is a generalization of the celebrated FitzHugh-Nagumo system, describing the nerve impulse propagation in axon. The modeling system is shown to possesses soliton-like solutions under certain restrictions on the parameters. The results of theoretical studies are backed by the direct numerical simulation.
}
\end{quote}

\begin{quote} \textbf{Keyword: }{\small 
models of active media; hyperbolic modification of the FitzHugh-Nagumo model; traveling waves; multidimensional dynamical systems; homoclinic solutions;  solitary wave solutions.
}
\end{quote}

\vspace{0.5 cm}

\section{Introduction}

  Studies of traveling wave (TW) solutions to nonlinear evolution equations attract attention of many researchers. Such interest is quite natural due to the fact that the TW solutions play an important role in the description of nonlinear phenomena in various fields of natural sciences, such as combustion and detonation \cite{ZeldBar,Zeld,SpinDet,CellStruct}, mathematical biology \cite{HodgHuxley,FitzHugh,Nagumo,A_Scott,Davydov,KPP}, nonlinear optics \cite{NSE} and hydrodynamics \cite{Dodd,Stoker,Tsunami}. A significant achievement of the  theory of nonlinear waves was the development of the theory of solitons at the end of the XX century \cite{Miura,Dodd,Zakharov}. Stability and particle-like properties of solitons, exposed  during interactions, are  often attributed to the Hamiltonian nature and complete integrability of the corresponding equations, manifested in presence of infinite hierarchy of conservation laws.  However, there is a large number of evolutionary equations of the dissipative type, which also have soliton-type solutions. In contrast to the Hamiltonian systems, dissipative models possess soliton-like solutions only for the selected values ​​of the parameters. Nevertheless, such solutions very often not only exhibit stability features, but also possess attracting properties \cite{Kamin_Rosenau,Barenblatt,KarchPudełko} and  are of particular interest as potential carriers of stable nonlinear perturbations in open dissipative systems.

The subject of interest of this work is study of the solitary wave solutions supported by the modification of the FitzHug-Nagumo equations \cite{FitzHugh,Nagumo} taking into account the effects of relaxation. For the FitzHug-Nagumo model, which is much simpler than the original Hodgkin-Huxley equations \cite{HodgHuxley}, it has become possible to show the existence of a moving pulse, prove its stability, and also confirm the existence of a threshold energy value below which the localized solutions become unstable. It should be noted that rigorous studies of the FitzHug-Nagumo model have been and remain a challenge to date. Therefore, much simple model is presented in the papers \cite{McKean,Wang,RinzK}. Simplification is achieved in this model (very often referred to as the McKean caricature on the FitzHug-Nagumo system)  by replacing the cubic nonlinear function, present in the FitzHug-Nagumo model, with a piecewise linear function. Later on it has ben proposed the following modification of the McKean system \cite{V_Lik}:
\begin{eqnarray}
\tau\,v_{tt}+v_t=v_{xx}+H(v-a)-v-w, \label{McGen1} \\
w_t=b\,v-d\,w. \label{McGen2}
\end{eqnarray}
A concept leading to the equation  with $\tau\,>\,0$ is presented in papers \cite{Joseph,Makar97,Kar03,DDMSV}. Equation (\ref{McGen1}) can be formally introduced if one changes in the balance equation for the variable $u$ the conventional Fick's Law
\[
J(t,x)=-K \nabla \, Q(t,\,x),
\]
stating the generalized thermodynamical flow-force relation, with the Cattaneo's generalization 
\[
\tau\,\frac{\partial}{\partial\,t}\,J(t,\,x)+J(t,\,x)=-K \nabla \, Q(t,\,x),
\]
which takes into account the effects of memory connected with the presence of internal structure on mesoscale.

In the present work we study the following system:
\begin{equation}\label{PDEq}
\begin{array}{c}
\tau  v_{tt}+v_t=v_{xx}+f(v)-w,\\
w_t=\epsilon (v-\gamma w ),\\
\end{array}
\end{equation}
where $f(v)=v(v-a)(1-v)$, $\gamma >0$, $ \tau >0$, $ \epsilon >0$.  Substantiation for this type of models was first proposed in papers \cite{Engelbrecht1992,Maugin1994}. 
Let us note that the global existence and uniqueness results to a class of systems more general than (\ref{PDEq}) have been presented recently in papers \cite{LusapaOM,LusapaRocky}.

Replacing a piecewise linear function with a function having the cubic nonlinearity greatly complicates the study, since it eliminates the possibility of constructing exact solutions of the soliton type. There are two main trends of research of soliton solutions in systems of type (\ref{PDEq}). In the concept based on the so called {\it slow-fast systems} approach \cite{Jones,CartSand}, the presence of a small parameter in the right side of the kinetic equation is essentially used. The approach we follow in this work is not directly related to the presence of a small parameter in the system, although the most important results are proved precisely for small values of the parameter $\epsilon$ and so far there is no reason to say that they can be transferred to a more general case. In this approach, initiated in \cite{Conley,Carpenter,Hastings_72,Hastings_76,Hastings_82}, the phase trajectories of a dynamical system associated with the initial system of PDEs are considered and their dependence on the parameters of the system are studied. The main problem we address in this work is the existence of homoclinic loops among the set of TW solutions, satisfying a multi-dimensional dynamical system. The importance of the homoclinic trajectories is due to the fact that they represent the nonlinear solitary waves. The proof of the existence of such trajectories, constituting the content of Section 2, is based on a number of additional statements presented in the form of lemmas and propositions. At the end of Section 2, the results of numerical experiments backing the analytical considerations are presented. In Section 3, the results obtained are summarized and the areas for further research are outlined.

\section{Existence of solitary wave solutions}

In what follows, we are interested in the traveling wave solutions  $v(t,\,x)=v(\xi),$ $w(t,\,x)=w(\xi),$ $ \xi=x+ct$, where $c>0$ is the velocity of the traveling wave, moving from right to left. Inserting these functions into (\ref{PDEq}),  we  get the  system
$$
\begin{array}{l}
\tau c^2\,v''= v'' - c v' +f(v)-w,\\
c w'= \epsilon (v-\gamma w ).\\
\end{array}
$$ 
Making the substitutions $\beta={1}/{(1-\tau c^2)},\,\,\delta=\epsilon/c $ and introducing new variable $u=v'$ we obtain the following dynamical system:
\begin{equation}\label{uklad2}
\begin{array}{l}
v'=u,\\
u'=  \beta\left[ c  u -  f(v)+ w\right],\\
 w'= \delta (v-\gamma w )\\
\end{array}
\end{equation}
(we assume further on that $\beta\,>\,0).$
Our aim is to show that, on certain restrictions on the parameters' values, the system (\ref{uklad2}) possesses homoclinic orbits, corresponding to the soliton-like TW solutions  of the initial system. We'll consider $a,\, \tau,  \, \gamma$ as auxiliary (fixed) parameters, whereas the parameters $\epsilon, $ and $ c$ as the main ones. Note that due to the assumption  regarding the sign of the parameter $\beta,$  the velocity $c$ cannot be arbitrarily large (which is unphysical), because it should satisfy the inequality $c^2\,<\,1/\tau$.  We'll also assume that 
\begin{equation}\label{restrgamma}
\frac{1}{\gamma}>\mathcal{N}=
\underset{v}{\max}\frac{f(v)}{v}=\frac{(a-1)^2}{4}.
\end{equation}
The above restriction assures that $(0,\,0,\,0)$ is the only stationary point of the system  (\ref{uklad2}).

\subsection{Local invariant manifolds of the origin}
Linearization matrix for the system (\ref{uklad2}) at the origin takes the form:
$$\hat A=
\left(\begin{array}{ccc}
 0  & 1 &  0 \\
 \beta a & \beta c  & \beta \\
 \delta & 0 & -\delta\gamma \\
\end{array}\right).
$$
The characteristic equation of the matrix $\hat A$  is as follows:
\begin{equation}\label{characteristic}
W(\lambda)=\frac{\beta \epsilon}{c}(1 + a \gamma ) + ( a \beta + \epsilon \gamma \beta ) \lambda +(c \beta -\frac{ \epsilon \gamma}{c} )\lambda^2- \lambda^3 =0.
\end{equation}
\begin{lem}
The matrix $\hat A$ has one positive eigenvalue $\lambda_1$ and a pair of eigenvalues $\lambda_{2,\,3}$  with negative real parts.
\end{lem}

\prf
The characteristic polynomial $W(\lambda)$ has always one positive real root $\lambda_1$ as it follows from the inequality $W(0)=\frac{\beta \epsilon}{c}(1 + a \gamma )>0$ and the asymptotic condition 
$\lim\limits_{\lambda \to\,+\infty }\,W(\lambda)=-\infty$. Using the Viete formulae
\begin{equation}\label{viet}
\left\{\begin{array}{c}
\lambda_1 \lambda_2 \lambda_3=  \frac{\beta \epsilon}{c}(1 + a \gamma ) >0,  \\
\lambda_1 \lambda_2 +\lambda_1 \lambda_3 +\lambda_3 \lambda_2 =  - \beta(a +\epsilon \gamma ) <0,  \\
\lambda_1+ \lambda_2 +\lambda_3= c \beta -\frac{ \epsilon \gamma}{c}, \\
\end{array}
\right.
\end{equation} 
one can easily check that $\Re\,\lambda_2,$   $\Re\,\lambda_3$ are negative.
\eprf

\begin{rem}
Let us note, that the eigenvector corresponding to $\lambda_j$  takes the form 
\begin{equation}\label{eigenv}
Y_j=\left(1,\,\lambda_j,\,\epsilon/(c\,\lambda_j+\epsilon\gamma) \right)^{tr},\,\,\, j=1,\,2,\,3.
\end{equation}
\end{rem}

Thus, under the restriction (\ref{restrgamma}), there exists a one-dimensional local invariant unstable manifold  $W_{loc}^u$ tangent to the eigenvector $Y_1$ at the origin and a two-dimensional local invariant stable manifold $W_{loc}^s$ tangent to the plane spanned by the vectors $Y_2,\,Y_3$. Note that $W_{loc}^u$ consist of two branches, one is tangent to $Y_1$ and pointed into the first octant (which is of interest to us), while the other is tangent to $-Y_1$.  

It is important to formulate the conditions assuring that $\lambda_{2,3}$ are  complex, since in this case the presence of a single homoclinic trajectory, under certain conditions, implies the presence of a countable set of homoclinic solutions in a small neighborhood of the parameter values for which a single bi-asymptotic trajectory does exist \cite{Feroe_83,Feroe_93,GonTurGas}. 

Let us note, that for values ​​of $\epsilon$  too close to zero, complex roots are absent. Indeed, the characteristic equation, that can be represented in this case as
\[
W(\lambda)=\lambda \left(  a\,\beta+ c\,\beta\lambda-\lambda^2 \right) =\,O(|\epsilon|),
\]
has only real roots when the $\epsilon$ tends to zero.

It is easy to see that the function $W(\lambda)$ has two extrema located at the points
\[
x_{\pm}=\frac{c\,\beta-\epsilon\,\gamma/c\,\pm\sqrt{\Delta}}{3}, \qquad 
\Delta=(c\,\beta-\epsilon\,\gamma/c)^2+ 3\,\beta(a+\epsilon\,\gamma) >0.
\]
Matrix $\hat A$ will have a pair of complex eigenvalues ​​if $W(x_{-})>0$, where $x_{-}$ is the point in which $W(\lambda)$ has the local minimum. 
This condition can be presented as follows:
\begin{equation}\label{cplxrt}
\eta+\frac{1}{9}\theta\,Y>\left(Y^2+3\,\theta   \right) 
\left[\sqrt{Y^2+3\,\theta}-\frac{2}{27} Y    \right],
\end{equation}
where
\[
\begin{array}{c}
Y=c\,\beta-\epsilon\,\gamma/c, \\
\eta=\frac{\beta\epsilon}{c}(1+a\,\gamma), \\
\theta=\beta(a+\epsilon\gamma).
\end{array}
\]

\subsection{Behavior of saddle separatrices in cases $\epsilon =0$ and $0<\epsilon \ll c \ll 1 $ }

In this subsection we use the condition $\epsilon =0$, allowing to separate the first two equations from the third one, having the trivial solution  $w = \it{const}.$ Arguments that will be given later on, allow us without loss of generality to restrict consideration to the case  $ w=0$. The remaining system then takes the  form
\begin{equation}\label{hamiltonowski}
\begin{array}{l}
v'=u,\\
u'=   \beta( c\,u- f(v)).\\
\end{array}
\end{equation}
For $c=0$ the system  (\ref{hamiltonowski}) can be presented in the Hamiltonian form with the Hamiltonian function $H= \frac{u^2}{2} +  \int f( v) dv $. This Hamiltonian corresponds to the so-called system with one degree of freedom. In most cases, such a system can be fully analyzed by qualitative methods (see e.g. \cite{AndrKhaikin}). The system (\ref{hamiltonowski}) has the following stationary points: $A=(0,\,0),$ $B=(a,\,0)$ and $C=(1,\,0).$ The first coordinate of each point corresponds to the extremal value of the potential energy
\begin{equation}\label{potenerg}
U_{p}(v)= \int f( v) dv=\frac{v^2}{12}\,\left[-6\,a+4(a+1)v-3\,v^2\right].
\end{equation}
The local minimum of (\ref{potenerg})  corresponds to the center, while the local maximum corresponds to the saddle point \cite{AndrKhaikin}. 
Analyzing the signs of the second derivatives of the function $U_{p}(v)$ at the corresponding points, one can conclude that for $0<a<1$  the points $A$ and $C$ are  saddles, while $B$ is a center. When $a>1,$ the point $A$ and $B$, in turn, are saddles, while the point $C$ is a center. We are looking for conditions assuring the existence of the trajectory doubly asymptotic to the saddle point $A$. Such  trajectory exists when either $a<1/2$ and $B$ is the center or when  $a>2$ and the center is located in the point $C$. 
It turns out that the case $a>2$ is not independent. Indeed, using in the case $a>2$ the scaling transformation
\[
T=a^2\,t,\quad X=a\,x, \quad \tau=\tilde \tau/a^2, \quad \tilde a=a^{-1}<1/2, 
\]
\[
\tilde v=v/a, \qquad \tilde w=s/a, \qquad \tilde \epsilon=\epsilon/a^2,
\]
one can write down the source system as follows:
\[
\begin{array}{c}
\tilde\tau  \tilde v_{TT}+\tilde v_T=\tilde v_{XX}+f(\tilde v)-\tilde w,\\
\tilde w_T=\tilde\epsilon (\tilde v-\gamma\tilde w ).\\
\end{array}
\]
Thus, turning to a solution dependent on the traveling wave variable $\xi=X-c\,T$ and putting in the factorized system $\tilde\epsilon=\tilde w=0$, we obtain the dynamical  system, up to the notation coinciding with (\ref{hamiltonowski}) in which $f(\tilde v)=\tilde v (\tilde v-1) \,(\tilde a -\tilde v)$ with $\tilde a<1/2.$  So we'll assume from now on that $0<a<1/2.$  Under the given assumption the stationary point $B=(a,\,0)$ is a center and, thus, there is an open set $U$ containing $B$, which is filled with the periodic trajectories. Since the $U_p(1)=(1-2\,a)/12\,>\,0,$ then the level line $U_p(v)=0$ corresponds to the homoclinic trajectory, formed by the separatrices of the saddle point $A=(0,\,0).$ Thus, at $c=\epsilon=w=0$ the system (\ref{hamiltonowski}) possesses the homoclinic solution. In what follows, we will need the information about the behavior of the separatrices of the saddle point $A$. We assume that $0<c \ll 1$ and denote the stable and unstable separatrices of the saddle point $A$ located in the right half-plane by $q_c^s(\xi)$ and $q_c^u(\xi),$ correspondingly.  The following statement holds true.

\begin{prop}
The saddle separatrice $q_c^u(\xi),$ directed towards the first quadrant
\begin{itemize}
\item
intersects the horizontal axis at some point $v_*$ such that $a<v_*<1;$
\item
 $q_c^u(\xi)=\left(v(\xi),\,u(\xi)   \right)$ tends to $(-\infty,\,-\infty)$ as $\xi\rightarrow \omega,$ where $0<\omega \leq +\infty.$
\end{itemize}
\end{prop}

\prf
The proof of the first item is based on the Melnikov theory \cite{Melnikov,GH}. Assuming that $c$ is small, we can present the system (\ref{hamiltonowski}), up to $O(c^2),$ in the following form
\begin{equation}\label{ham_pert}
\left(\begin{array}{c} v \\ u \end{array} \right)^\prime =F+c\,G,
\end{equation} 
where $F=(\partial\,H/\partial\,u,\,-\partial\,H/\partial\,v)^{tr}=(u,\,-f(v))^{tr},$   $G=(0,\,u)^{tr}.$  For $c>0$ the stationary point $B$ turns into unstable focus, and stable and unstable separatrices of the saddle $A$ do not form a closed loop any more. We want to trace what happens with the stable and unstable separatrices when $0<c$ is small. Let us denote the point at which the homoclinic loop corresponding to $c=0$ intersects the horizontal axis by $p.$ For $0<c\ll 1$ the stable and unstable separatrices will be located in the  neighborhood of the homoclinic curve. We denote by $q_c^s(0)$ and $q_c^u(0)$   the points at which the stable and unstable separatrices intersect the horizontal axis (see Fig. 1).
\begin{figure}[th]
\begin{center}
\includegraphics[totalheight=2 in]{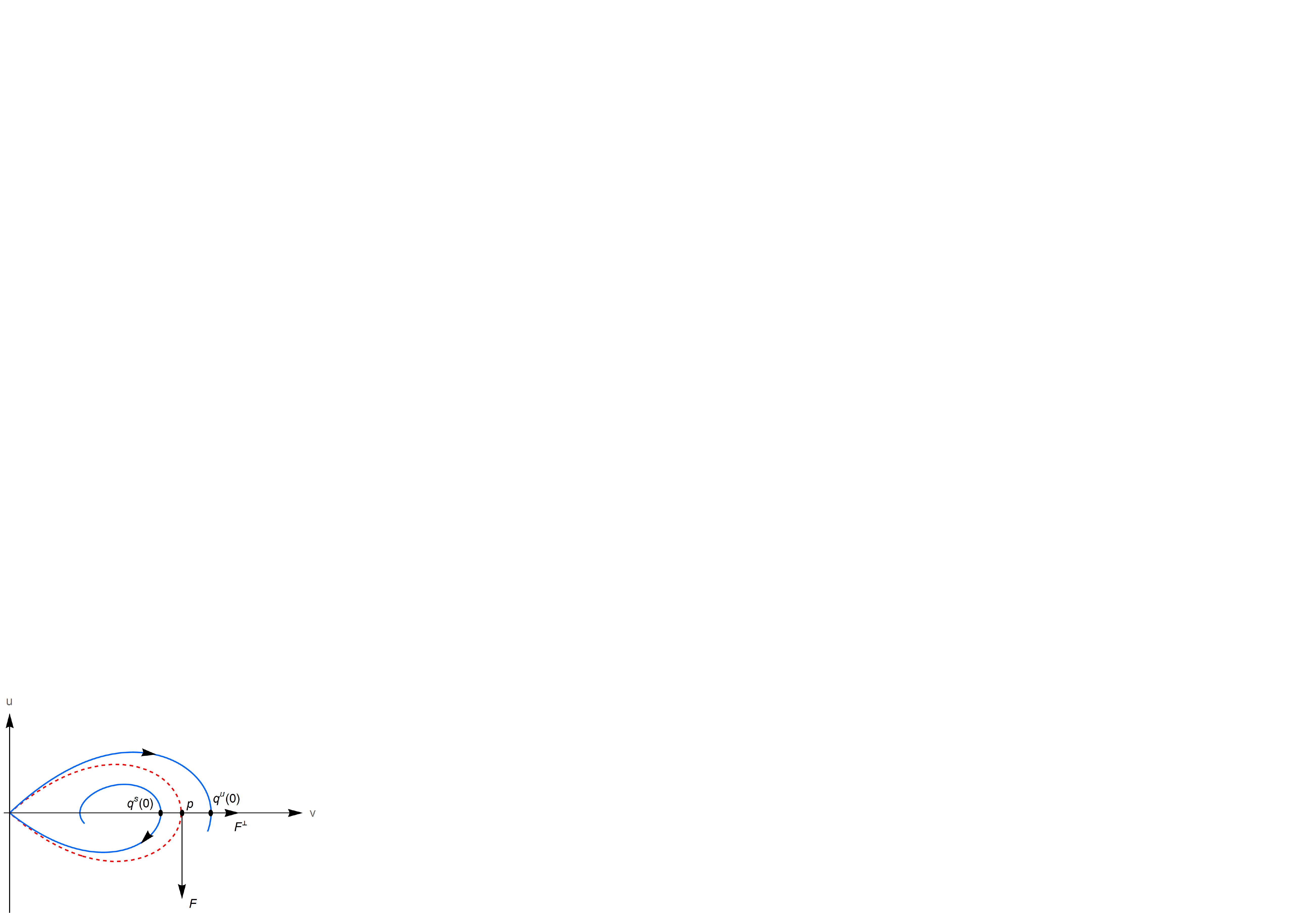}
\caption{The homoclinic loop corresponding to the case $c=0$ (dashed line) versus the  separatrices corresponding to $0<c\ll 1$}\label{GSV:phasePort}
\end{center}
\end{figure}%Figure1
Up to $O(c^2),$ the projection of the vector $q^u_c(0)-q^s_c(0)$ onto the vector $F^{\bot}=(f(v),\,u)^{tr}$ is given by the Melnikov integral {\cite{Melnikov,GH}}
		\[
		M=c\int_{-\infty}^{+\infty}{F^{\bot}\,\cdot\,G}\,d\xi=c\int_{-\infty}^{+\infty}{u^2(\xi)\,d\xi}>0. 
		\]
Thus the stable and unstable separatrices in the case $c>0$ form the configuration shown on Fig.~\ref{GSV:phasePort}. 

Further behavior of the saddle separatrice $q_c^u(\xi)$ is following. After the intersection of the horizontal axis, it enters the fourth quadrant and its coordinate $u(\xi)$ remains negative further on, since the trajectory is separated from the upper half-plane by the separatrices of the saddle points $A$ and $C$. Thus the coordinate $v(\xi)$ decreases as $\xi$ grows. The coordinate $u(\xi)$, in turn, decreases when $v(\xi)>a$ and increases when $0<v(\xi)<a,$ remaining negative. It becomes decreasing function again when $v(\xi)$ is negative, and from this instant $q^u_c(\xi)$ monotonically tends to $(-\infty,\,-\infty).$   \eprf

Let us analyze the behavior of solutions for nonzero $\epsilon$ and $c$ satisfying the conditions $0<\epsilon \ll c \ll 1.$ Up to the term of the order $O(c^2)$, the system  (\ref{uklad2}) in this case can be presented as follows
\begin{equation}\label{uklad2A}
\begin{array}{l}
v'=u,\\
u'=   c \, u -  f(v)+ w,\\
 w'= \delta (v-\gamma w ),\\
\end{array}
\end{equation}
where $\delta=\epsilon/c\ll 1.$ We are interested in the behavior of the trajectory $q^u_{\epsilon\,c}(\xi)=\left(v^u_{\epsilon\,c}(\xi),\,u^u_{\epsilon\,c}(\xi),\,w^u_{\epsilon\,c}(\xi)   \right)$ being the three-dimensional deformation of the trajectory $q^u_c(\xi)$ and satisfying the condition $\lim\limits_{\xi \to -\infty}q_{\epsilon\,c}^u(\xi)=0.$ The analysis of the linearization of the system  (\ref{uklad2A}) shows that such deformation does exist. Without the loss of generality, we can assume that $v^u_{\epsilon\,c}(0)=a,$ $u^u_{\epsilon\,c}(0)>0,$ $\left(v^u_{\epsilon\,c}\right)^\prime(\xi)|_{(-\infty,\,0)}>0$, and $\left(u^u_{\epsilon\,c}\right)^\prime(\xi)|_{(-\infty,\,0)}>0.$ 

Presenting the solution to the third equation of the system (\ref{uklad2A}) in the form $w=\delta\,w_1+O(\delta^2),$ we obtain with the specified accuracy the following representation:
\begin{equation}\label{W_approx}
w_1(\xi)=\int_{-\infty}^\xi{v(y)\,d\,y,}
\end{equation}
where $v(z)$ is the first coordinate of the unstable saddle separatrice of the stationary point $A$ of the system (\ref{hamiltonowski}).  
For $\delta \ll 1,$ $v^u_{\epsilon\,c}(\xi)$ still dominates the behavior of the third variable and therefore $w^u_{\epsilon\,c}$ in the r.h.s of the second equation does not influence the qualitative behavior neither the variable $u^u_{\epsilon\,c}(\xi)$ nor the variable $v^u_{\epsilon\,c}(\xi),$ which becomes negative and monotonically decreasing from some instant. And
when the function  $|v^u_{\epsilon\,c}(\xi)|$ becomes large enough, all three functions monotonically tend to $-\infty.$ Let us formulate the result obtained as follows.
\begin{cor}
There exist an open set $\Delta$ in the space of the parameters $(c,\,\epsilon)$, placed at the first quadrant and adjacent to the horizontal axis such that for all $(c,\,\epsilon)\,\in\,\Delta$ the phase trajectory
$q^u_{\epsilon\,c}(\xi)$
% satisfying the conditions $v^u_{\epsilon\,c}(0)=a,$ %$u^u_{\epsilon\,c}(0)>0,$  $w^u_{\epsilon\,c}(0)>0,$ %$\left(v^u_{\epsilon\,c}\right)^\prime(\xi)|_{-\infty,\,0}>0$, %$\left(u^u_{\epsilon\,c}\right)^\prime(\xi)|_{(-\infty,\,0)}>0,$ %$\left(w^u_{\epsilon\,c}\right)^\prime(\xi)|_{(-\infty,\,0)}>0,$ %$\lim\limits_{\xi \to -\infty}q^u_{\epsilon\,c}(\xi)=0$  
satisfies the condition
$
\lim\limits_{\xi \to +\infty}q^u_{\epsilon\,c}(\xi)=(-\infty,\,-\infty,\,-\infty).
$
\end{cor}

\subsection{The sets positively invariant  with respect to the phase flow of the system (\ref{uklad2})}

In this subsection we will prove the following lemma:
\begin{lem}
The sets:
$$ 
\begin{array}{c}
E^+=\lbrace (v,u,w): \, u>0, v>1, w'>0, u'>0 \rbrace, \\
E^-=\lbrace (v,u,w): \, v<0, u<0, w'<0, u'<0 \rbrace\\
\end{array}
$$
are positively invariant with respect to the phase flow $\phi_t$ generated by the dynamical system (\ref{uklad2}).
\end{lem}
\prf
In the proof below, as well as in the proofs of the subsequent assertions, we mainly follow the plan drawn in Ref. \cite{Hastings_82}.
Thus, suppose that $E^+$ is not positively invariant with respect to the $\phi_t$, and the solution $q(\xi)=\left(v(\xi),\,u(\xi),\,w(\xi)\right)$ satisfying $q(0) \in E^+$ leaves the set $E^+$ for the first time at $\xi_1>0,$ so one of the features characterizing this set fails. To begin with, let us observe that the equality $v(\xi_{1})=1$ cannot be true, since $v(\cdot)$ is growing on the segment $(0,\,\xi_1)$ and $v(0)>1.$ For the same reason $u(\xi_1)$ cannot be equal to zero. Now let us address the function $w'(\cdot).$ The relation $w'|_{(0,\,\xi_1)}>0,$ together with the supposition $w'(\xi_1)=0,$ imply the inequality $w''(\xi_1)\leq 0,$ but $w''(\xi_1)=\frac{\epsilon}{c}u(\xi_1)>0,$ hence we get the contradiction. Now, let us consider $u'(\xi).$ In accordance with the above assumptions  
$u'(0)=\beta\left\{c\, u(0)+w(0)-f[v(0)]\right\}>0$ and since $u(\cdot),$ $w(\cdot),$ and $v(\cdot)$ are growing functions on the segment $(0,\,\xi_1),$  and so is $-f[v(\cdot)],$ then $u'(\xi_1)$ cannot be zero as well. The positive invariance of the set $E^-$ is shown just in the same manner.    
\eprf

\subsection{Asymptotic behavior  of the unstable invariant manifold}

We still assume that $q^u_{c,\epsilon}(\xi)=\left(v^u_{c,\epsilon}(\xi),\,u^u_{c,\epsilon}(\xi),\,w^u_{c,\epsilon}(\xi)\right)$ is the unstable invariant manifold of the stationary point $(0,\,0,\,0),$ corresponding to the given values of the parameters $c,\,\epsilon$. however, for simplicity, from now on we will omit the superscript. Without the loss of generality we'll also assume that  $v_{c,\epsilon}(0)=a$ and $u_{c,\epsilon}(\xi) >0$ when $\xi\,\leq\,0.$ Let us define  the following subsets of the set $ \Omega = \lbrace ( c, \epsilon) \, |\,0\,< c\,<1/\sqrt{\tau},\,\, \epsilon \geq 0 \rbrace$:
$$
\begin{array}{lll}
\Omega_1& = &\lbrace ( c, \epsilon) \in \Omega \, | \,  q_{c,\epsilon}(\xi) \, \mathrm{is} \,\, \mathrm{bounded} \rbrace,\\
\Omega_2& = &\lbrace ( c, \epsilon) \in \Omega \, | \,\,\lim\limits_{\xi \to +\omega_0} (v_{c,\epsilon} (\xi), \, u_{c,\epsilon} (\xi)) = (+\infty ,+\infty ) \rbrace,\\
\Omega_3 &= &\lbrace ( c, \epsilon) \in \Omega \, | \, \,\lim\limits_{\xi \to +\omega_1} (v_{c,\epsilon} (\xi),\, u_{c,\epsilon} (\xi)) = (-\infty ,-\infty ) \rbrace,\\
\end{array}
$$
where $0\,<\,\omega_i\, \leq\, +\infty,\,\,i=0,\,1.$
We will show that there is no other possible behavior of the trajectory $q_{c,\epsilon}(\xi)$ differing from that presented above. 
\begin{lem} The following statement is true:
$
\Omega= \Omega_1 \cup \Omega_2 \cup \Omega_3.
$
\end{lem}
\prf
We construct such a rectangle $A$ in the plane $(v,\,w)$ that  there will be only three possibilities:
\begin{itemize}
\item solution  is bounded by this rectangle,
\item solution leaves the rectangle and after that it enters $E^+$, which implies that $ q_{c,\epsilon} (\xi) \rightarrow (+\infty,\,+\infty,\, +\infty )$,
\item solution leaves the rectangle and after that it enters $E^-$, which implies that $ q_{c,\epsilon} (\xi) \rightarrow (-\infty ,\,-\infty, \,-\infty ).$ 
\end{itemize}
We define a rectangle
$$
A= \lbrace (v,w)\, :\,\,  v_1 \leq v \leq v_2, \,  w_1 \leq w \leq w_2 \rbrace , 
$$
assuming that $v_2>1$, $ w_2= \frac{v_2}{\gamma}$, and  $v_1$ is such that
$$%\begin{array}{cc}
f(v_1)=\frac{v_2}{\gamma}, \qquad w_1= \frac{v_1}{\gamma}.
%\end{array}
$$
It can be shown that choosing $v_2>1$ sufficiently large, we will get the inequality $w_1>f(v_2).$
\begin{figure}[th]
\centering
\includegraphics[scale=0.8]{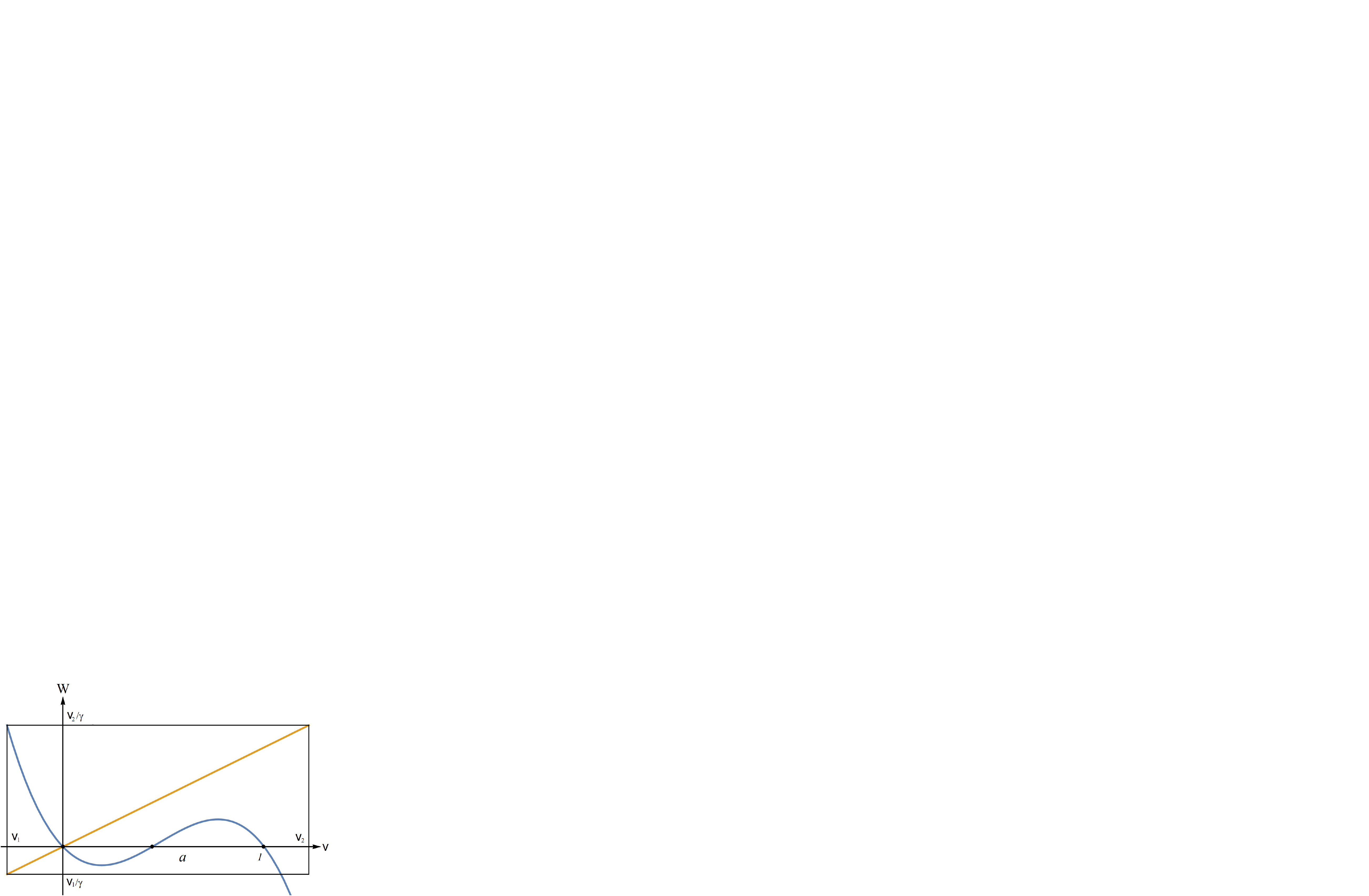}
%prostokat.eps
\caption{Rectangle $A$}\label{Rectangle}
\end{figure}%Figure2
Then for
 $w<\frac{v}{\gamma},\,\,$ 
$w'= \frac{\epsilon}{c} (v-\gamma w ) >0$. In addition, we have the inequalities 
$u'=\beta[ c u - f(v_2) +\frac{v_1}{\gamma}] >0$, 
and $v'=u>0$.
This implies that $ q_{c, \epsilon}(\xi) \in E^+$ and hence tends to $( +\infty, +\infty, +\infty)$ for $t \rightarrow \infty,$  so the pair $(c,\epsilon) \in \Omega_2$.

Just in the similar way, one can prove that if $w>\frac{v}{\gamma}$, then the phase trajectory approaches the left boundary of the rectangle $A$ and ultimately falls into the set $E^-$, so the corresponding pair $(c,\,\epsilon)$ belongs to the set $\Omega_3$. 
 
To complete the proof of this part, it suffices to note that   
the phase trajectory cannot leave the rectangle  by its top or bottom. But it is quite evident, since at the points belonging to the upper and lower borders the vector field is directed inward the rectangle $A$. And if the solution $q_{c, \epsilon}$ does not leave the set $A$, then the pair $(c,\epsilon)$ belongs to $ \Omega_1$ and since there is no other choice, the statement is completely proved.
\eprf

\subsection{Insight into the structure of the subsets $\Omega_2$ }
In this subsection, the geometry of subset of the parameters $(c,\,\epsilon)\,\in\,\Omega$ for which all the components of the vector-function $ q^u_{c,\epsilon}(\xi)$ go to $+\infty$  will be highlighted.
\begin{theo}\label{L4}
There exists values  $c_1>0$ and  $\epsilon_1>0$ such that: 
$$\lbrace (c,\epsilon):\,\,c_1<c<1/\sqrt{\tau},\,\, \mathrm{or}\,\, \epsilon > \epsilon_1 \rbrace  \subset \Omega_2. $$\label{Tw1}
\end{theo}
\prf
Without the loss of generality, we assume that  $v(0)=a>0,$ $u(0)>0$ and $w(0)>0.$ The positivity of all components of the vector $q(0)$ appears from the fact that for $\xi\,\ll\,-1$ the phase trajectory is close to the eigenvector $Y_1=\left(1,\,\lambda_1,\,\epsilon/(c\,\lambda_1+\epsilon\,\gamma)\right)^{tr}$ and for $\xi\,\in\,(-\infty,\,0)$ all three components are non-decreasing. Indeed, for  $\xi\,\ll\,-1$
${w(\xi)}/{v(\xi)}\approx {\epsilon}/(c\,\lambda_1+\epsilon\,\gamma) 
$
and if the r.h.s is less that $\gamma^{-1}$ then the projection of $q(\xi)$ onto the plane $(v,\,w)$ lies below the line $w=v/\gamma.$ But this requirement is equivalent to the inequality $\epsilon\,\gamma<c\,\lambda_1+\epsilon\,\gamma,$ which is true because all the parameters are positive. For $\xi\,\in\,(-\infty,\,0)$ the r.h.s. of the third equation of the system (\ref{uklad2}) remains non-negative because at the instant when the projection of the trajectory onto the plane $(v,\,w)$ approaches the line $w=v/\gamma,$ $v(\xi)$ is positive and the  projection cannot cross the line $w=v/\gamma$. This, in turn, implies that all components of the vector $q(\xi)$ for $\xi\,\in\,(-\infty,\,0)$ are positive, and besides $v(0)-\gamma\,w(0) \,\geq\,0.$ 

Further, as long as $v(\xi)$ is an increasing function, we can assume that
$u(t)=U(v(t))$, $w(t)=W(v(t))$. Then :
\begin{equation}\label{UW_v}
\begin{split}%{c}
\frac{d\,U}{d\,v} &= \beta \left(c   + \frac{W-f(v)}{U}\right), \\ 
\frac{d\,W}{d\,v} &= \delta\frac{ V  - \gamma W}{U}.
\end{split}
\end{equation}
Suppose that $U[v(0)]=U(a)>f_{max}/c,$ where $f_{max}= \underset{v\in [0,1]}{\sup}f(v)$=$\underset{v\in [a,1]}{\sup}f(v)$. The r.h.s. of the second equation of the system (\ref{UW_v}) is non-negative until $U$ is positive. But $U(a)>0,$ and under the above supposition, 
\[
\frac{d\,U}{d\,v}>\beta\,c\,\left(1-\frac{f(v)}{f_{max}}   \right)\,>\,0,
\]   
so $U(v)$  is growing, $W(v)$ is non-decreasing as $v>a,$ and $q(\xi)$ attains the set $E^+$.

Now let us  assume that $0<U(a)\,\leq\,f_{max}/c,$ and $W(a)>f_{max}.$ Then we get the estimation
\[
\frac{d\,U}{d\,v}>\beta\left(c+\frac{W-f_{max}}{U}   \right)\,>\beta\,c>0,
\]   
 and $q(\xi)$ attains the set $E^+$. Now it is necessary to find the conditions assuring that the inequality  $W(a)>f_{max}$ is fulfilled.  Under the assumption $U(a)\,\leq\,f_{max}/c,$ the inequality 
\[
\frac{d\,W}{d\,v}>\frac{\delta\,c}{f_{max}}(v-\gamma\,W)
\]
takes place on the segment $v\,\in\,(0,\,a).$ Applying the substitution 
$W(v)=A(v)\,e^{-\rho\,v},\,\,\rho=\epsilon\,\gamma/f_{max},$ we get the inequality 
\[
A^\prime(v)>\frac{\delta\,c}{f_{max}}\,v\,e^{\rho\,v}, 
\]
which, after the integration w.r.t. $v$ on the segment $(0,\,a)$ takes the form
\[
A(a)>\frac{\theta}{\rho^2}\left[\left(a\,\rho-1\right)\,e^{\rho a}+1\right],
\]
where $\theta=\epsilon/f_{max}.$ 
From this we get the inequality
\[
W(a)>\frac{\theta}{\rho^2}\left(e^{-\rho a}+\rho\,a-1   \right)=
\mu \frac{\epsilon\, a^2}{2\,f_{max}},
\]
where $0<\mu<1.$ So, if $\epsilon>\epsilon_1=2 \,f_{max}/(\mu\,a^2),$ then, regardless of the value of $c>0$, $W(a)>f_{max}$.

Now let us estimate $c_1$.  The first equation of the system (\ref{UW_v}) can be rewritten in the form
\[
\frac{1}{2}\frac{d}{d\,v}U^2(v)=\beta\left[c\,U+W-f(v)   \right],
\]
from which appears the inequality
\[
U(a)\,\geq\,H(a):=\left[-2\,\beta\int_0^a{f(v)\,d\,v}   \right]^{1/2}.
\]
If $c>f_{max}/H(a)$, then 
\[
\beta \left[c\,U+W-f(v)\right]>\beta\,W>0.
\]
To complete the proof, we just show that the inequalities  $c>f_{max}/H(a)$ and $c<1/\sqrt{\tau}$ are compatible. The first one is equivalent to
\[
\frac{c^2}{1-\tau\,c^2}>\frac{f^2_{max}}{-2\,\int_0^a{f(v)\,d\,v}}:=\sigma>0,
\]
or 
\[
\frac{\sigma}{1+\tau\sigma}<c^2<\frac{1}{\tau}\equiv \frac{\sigma}{\tau\sigma},
\]
so the inequalities are compatible and the statement is completely proved. 
\eprf

Now, let us show that the following assertion is true.

\begin{lem}
If the parameters of the system (\ref{uklad2}) belong to the set   
\begin{equation}\label{ineqtriv}
\Phi=\left\{(c,\,\epsilon)\,|\,\mathcal{N}=\frac{\beta (a-1)^2}{4}< \frac{\epsilon}{c^2}<\frac{\beta}{\gamma}  \right\},
\end{equation} 
then the only bounded solution possible  is the trivial solution $q_{c,\epsilon}(\xi)=0$. \label{lem2}
\end{lem}
\prf
Suppose that under the above conditions there exists a nontrivial bounded solution $q^u_{c,\epsilon}(t).$ 
By analogy with \cite{Conley,Hastings_82}, we consider the function 
$$G(v,u,w)=u^2+ 2\beta \int_0^v f(z)dz+(c\beta)^2\,v^2+\beta^2\left(\frac{c^2 }{\epsilon}-\frac{\gamma}{\beta}  \right)\,w^2-2\,\beta\,v(w +c\,u)$$
which, under certain conditions, is monotonically decreasing on the solutions of the system (\ref{uklad2}). For simplicity we drop the subscripts (superscripts) in what follows.  

Differentiating the function $G[q(\xi)]$, and taking into account (\ref{uklad2}), we obtain:
\[
\begin{array}{c}
\frac{d}{d\,\xi}G[q(\xi)]=-2\,\beta  \,c \,v^2\left(\frac{\epsilon}{c^2}-\beta\frac{f(v)}{v}\right)-2\,\gamma\delta\beta^2\,w^2\left(\frac{c^2}{\epsilon}-\frac{\gamma}{\beta}   \right).
\end{array}
\]
In view of the assumption (\ref{restrgamma}), ${d}\,G[q(\xi)]/{d\,\xi}<0$ if the parameters $c,\,\epsilon$ belong to the set $\Phi$ and $v^2+w^2 \neq 0.$  The existence of such function suggests that $G[q(\xi)]$ should tend monotonically to a finite value differing from zero as $\xi$ tends to $+\infty$, but this is impossible, since the origin is the only stationary point of the system (\ref{uklad2}).       
 \eprf
 
\begin{lem}
The sets $ \Omega_2$ i $\Omega_3$ are relatively open in $\Omega$.
\end{lem}
\noindent
The statements appears from the fact that  the sets $ E^+$ and $E^-$ are open and positively invariant while the solutions of the system (\ref{uklad2}) continuously depend on the parameters.
%
%{}

\begin{figure}[th]
\centering
\includegraphics[scale=0.6]{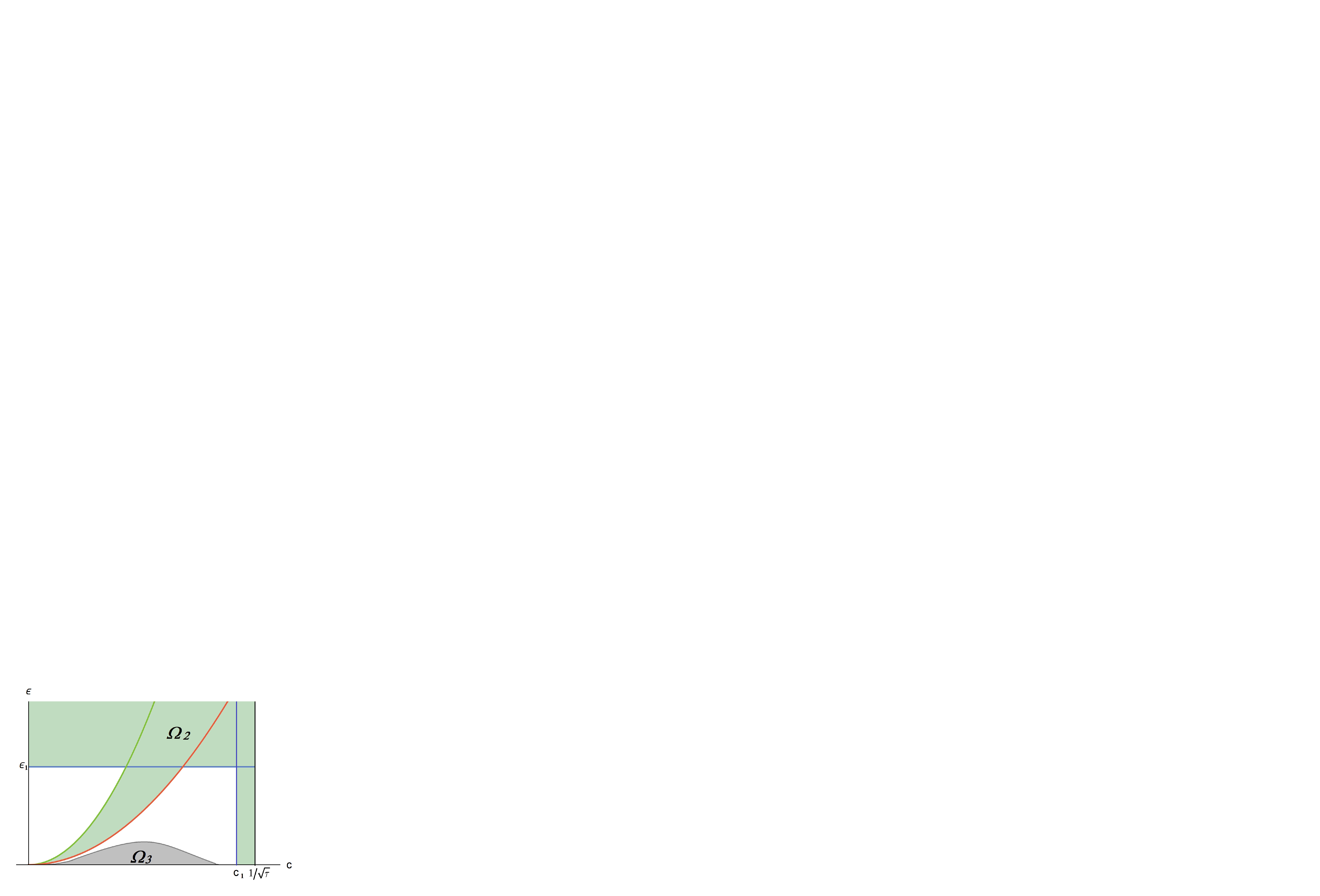}
\caption{The geometry of the sets $\Omega_2$,  $\Omega_3$}\label{OM23}
\end{figure}

\begin{cor} 
The set $ \Phi$ belongs to  $ \Omega_2$.
\end{cor}
\prf
All the solutions $q_{c,\,\epsilon}(\xi)$ corresponding to $(c,\,\epsilon)\in\,\Phi\,$ are unbounded. On the other hand, the set $\Phi$ is open and has nonempty intersection with the set $\Omega_2$ (see Fig.~\ref{OM23}). Since $\Omega_2$ and $\Omega_3$  are relatively open and disjoint, then $\Phi\in\Omega_2.$ 
\eprf
%Figure3

It is seen from the geometry of the open sets $\Omega_2$ and  $\Omega_3$ (Fig.~\ref{OM23}), that there should exist two subsets of the set $\Omega_1$, lying between them. One of these subsets is adjacent to the origin, while the second one lies closer to the line  $c=c_1.$ It remains to show that among the elements of these subsets there are pairs $(c,\,\epsilon)$ corresponding to the orbits bi-asymptotic to the origin.

\subsection{Solutions  corresponding to the solitary waves}

Let us consider the set

\[ \Sigma = \bigl\{ (v,u,w) |\, v \geq v_{min},\, u=0,\, \,\mathrm{or} \,\, v=v_{min},\,  u < 0 \bigr\}, 
\] 
where $v_{min}$ is the point of a local minimum of the function $f(v)$ on the segment $(0,\,1),$ and
\[ \Lambda = \biggl\{ \begin{array}{l} ( c, \epsilon )\,\in\,\Omega\,: \, \mathrm{solution} \,\, q_{c,\,\epsilon}(\xi)\,\, \mathrm{intersect}\, \, \Sigma \, \,\mathrm{exactly} \, \mathrm{two}\, \mathrm{times}  \\ 
 \mathrm{and} \, \mathrm{after} \, \mathrm{that} \, \mathrm{does} \, \mathrm{not} \, \mathrm{intersects} \,\mathrm{the}\, \mathrm{region}  \, v \,\geq\, v_{min} \end{array} \biggr\}. \]
\begin{figure}[h]
\centering
\includegraphics[scale=0.4]{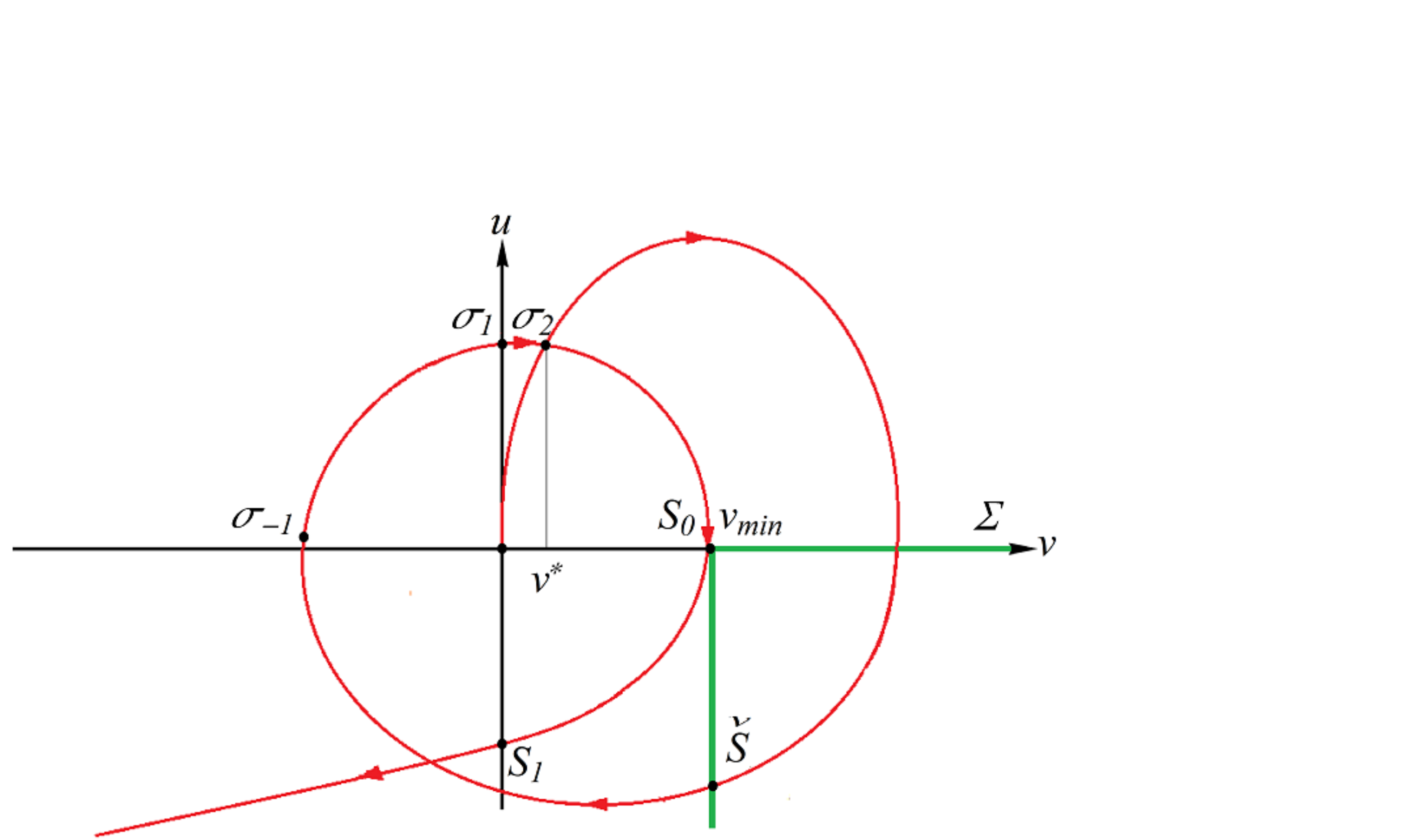}
\caption{Projection of the phase trajectory $q_{c,\,\epsilon}$ onto the plane $(v,\,u)$}\label{vupr}
\end{figure}%figure4
\newline
It follows from the definition of the set $\Lambda$ that 
$\Omega_2 \cap \bar{ \Lambda} = \emptyset$. On the other hand, the arguments following the proof of the Proposition 1 imply that the set $\Lambda$ contains a non-empty open subset of points belonging to $\Omega_3.$ And now we are going to prove the following assertion.

\begin{theo}
If  $\epsilon>0$ is sufficiently small, then  $( c, \epsilon) \in \Omega $ does not belong to the set $ \Omega_3 \cap \partial  \Lambda.$  \label{theo2}
\end{theo}

\prf
The proof of this theorem is based on the ideas underlying the proof of the Lemma 9 of the paper \cite{Hastings_82}, so we try to adhere to the notation that was adopted in this work. 

Let us suppose the opposite, namely, that to every pair $(c,\,\epsilon)$ belonging to the set $ \Omega_3 \cap \partial  \Lambda$ corresponds the orbit $q_{c,\,\epsilon}(\xi)$ with the following properties:
\begin{itemize}

\item

the orbit  $q_{c,\,\epsilon}(\xi)$ starts from the origin and points to the first octant as $\xi \ll -1$. We can assume without the loss of generality that $v^\prime(\xi)$ and $u^\prime(\xi)$ are positive on the interval $(-\infty,\,0)$, and $v(0)=a$;

\item

at some value of the argument, say $\xi=s_{-}>0,$ the orbit crosses the set $\Sigma\,$ for the first time, intersecting it at a point belonging to the plane $v\,>\,v_{min},$ $u=0,$ and next at  $\xi=\,$ {\it \v{s}} $\,>s_{-}$ intersects $\Sigma$ for the second time at a point belonging to the plane $v\,=\,v_{min},$ $u\,<\,0;$

\item

before crossing the plane $v=0$ and going to infinity  (suppose that such intersection take place at $\xi=s_1$), the orbit must touch the set $\Sigma,$ say, at $\xi=s_0>\,$ {\it \v{s}} (otherwise it does not belong to the set $\partial\,\Lambda$). Analysis of the first equation of the system (\ref{uklad2}) tells us that the touch point must be located at the intersection of planes $\left\{v\,=\,v_{min}, \, u\,<\,0 \right\}$ and $\left\{v\,>\,v_{min}, \, u=0 \right\}.$   

\end{itemize}
Projection  of an orbit  $q_{c,\,\epsilon}(\xi)$ on the plane $(v,\,u)$ is schematically represented in Fig. \ref{vupr}. It is obvious that, since  the orbit $q^u_{c,\,\epsilon}(\xi)$ is tangent to the set $\Sigma,$ then there exists a number $\mu>0$ such that $u(\xi)>0$ at $(s_0-\mu,\,s_0),$ $u(\xi)<0$ at $(s_0,\,\,s_0+\mu)$ and $u(s_0)=0.$  Looking at the second equation of the system (\ref{uklad2}), we easily conclude that 
\begin{equation}\label{ws0}
w(s_0) \,\leq\,f(v_{min})<0.
\end{equation}
It will be shown below that, for sufficiently small $\epsilon$, there does not exist the trajectory $q_{c,\,\epsilon}$ with $(c,\,\epsilon)\,\in \, \Omega_3 \cap \partial  \Lambda,$ which is characterized by the relations 
\[
v(s_0)=v_{min}, \quad u(s_0)=0, \quad  \quad u^\prime(s_0)\leq 0.
\]
So, let $\xi=\sigma_1$ be the closest to $s_0$ point such that $v(\sigma_1)=0$ and $\sigma_1<s_0.$ Let us denote by $\sigma_0<s_0$ the value of the argument nearest to $s_0$ from the left, at which $w(\xi)$ attains zero (it is easily seen that $\sigma_0\,\leq\,\sigma_1$). Thus, the function $v(\xi)$ increases on the interval $(\sigma_1,\,s_0)$, while the function $w(\xi)$ decreases on the interval $(\sigma_0,\,s_0).$  Let us choose a point $v^{*}\,\in\,(0,\,v_{min})$ and let $\sigma_2$ be the value of the argument closest to $s_0$ from the left, at which $v(\sigma_2)=v^{*}.$ Thus we have the relations {\it \v{s}}$\,<\sigma_0 \leq \sigma_1 < \sigma_2 < s_0 <s_1$ (the points of the orbit $q_{c,\,\epsilon}$ corresponding to these values of the arguments are shown schematically in Fig. \ref{vupr}). Next, it is seen from the third equation of the system (\ref{uklad2}), that the function $w(\xi)$ increases at the segment $(\sigma_1,\,s_0),$ remaining negative. Therefore the inequality $w(\xi)-f(v_{min})<0$, stated above for $\xi=s_0$, is valid for $\xi \in (\sigma_1,\,s_0).$ We also can state that $u(\xi)>0$ and $u^\prime(\xi)<0$ for $\xi \in (\sigma_1,\,s_0).$ Indeed, if there is a point $\sigma \in   (\sigma_1,\,s_0)$ such that $u^\prime(\sigma)=0,$ then taking the derivative of the second equation of the system (\ref{uklad2}) we get
\[
u^{\prime\prime}=\beta \left\{\delta \left[v(\sigma)-\gamma w(\sigma)\right]-f^\prime[v(\sigma)]\,u(\sigma)   \right\}>0,
\]
which leads to the contradiction. 

Next, we are going to choose the parameter $\alpha>0$ so that the following inequality takes place 
\begin{equation}\label{alfa}
\beta \left[ c+\frac{f_{min}-f(v_{*})}{\alpha}  \right]\,v^{*}+\alpha<0,
\end{equation} 
where $f_{min}=f(v_{min})$ (note that $\kappa=f(v_{*})-f_{min}>0$). We want to choose the parameter $\alpha$ so that the inequality (\ref{alfa}) be satisfied for any value of $c$ from the interval $(0,\, c_1).$ Such a choice is possible, since the roots of the quadratic equation
\[
\alpha^2+\beta\,c\,v^{*}\,\alpha-\kappa\,\beta\,v^{*}=0
\]
have different signs for any $c\,\in\,[0,\,c_1].$ We are going to analyze two possible cases. 

Suppose first that $0<u(\sigma_1)<\alpha.$ We can again use the monotony of $v$ on the segment $(\sigma_1,\,\sigma_2)$ and stated above fact that $u(\xi)$ is positive and decreasing on this interval. This leads to inequality
\[
\frac{d\,U}{d\,v}\,\leq\,\beta\left\{c+\frac{f_{min}-f(v^{*})}{\alpha}   \right\}.
\]
Integrating this inequality with respect to the variable $v$ within the interval $(\sigma_1,\,\sigma_2)$ we get the inequality
\[
U(v^{*})\,\leq\,\beta \left[ c+\frac{f_{min}-f(v_{*})}{\alpha}  \right]\,v^{*}+\alpha<0.
\]
But this contradicts the previously obtained inequality.

Now suppose that  $u(\sigma_1)\,\geq\,\alpha.$ Since the function $u(\xi)$ is positive and increasing on the segment $(-\infty,\,0)$, while  on the segment $(\sigma_1,\,s_0)$ it is decreasing function, then there are points where this function changes its sign.  Let $\sigma_{-1}$ be the largest value of $\xi<\sigma_1$ where $u^\prime=0.$ From this appears that $u^{\prime\prime}(\sigma_{-1})\,\leq\,0.$ Evaluating the behavior of $u^{\prime\prime}$, we will show that for sufficiently small $\epsilon$ this is not true. Since  $u(\sigma_1)\,\geq\,\alpha,$ then we get at $\xi=\sigma_{-1}$ the estimation
\[
u^{\prime\prime}=\beta\left\{\delta\,\left(v-\gamma\,w\right)-f^\prime\,(v)\,u   \right\}\,\geq\,  \beta\,\left\{\delta\,\left(v-\gamma\,w\right) + a\,\alpha \right\}.
\]
Since the trajectory leaves the origin pointing towards the first octant, its projection onto the plane $(v,\,w)$ should  still be in the rectangle $A,$ shown in Fig. \ref{Rectangle} when $\xi=\sigma_{-1}$ (otherwise the phase trajectory can no more touch the set $\Sigma$). For this rectangle the following estimation holds:
\[
|v-\gamma\,w|\,\leq\,2\,\tilde{V},
\]
where $\tilde{V}=\max\{|v_1|,\,v_2\}.$ For the parameter $\delta=\epsilon/c$ we get the estimation
\[
\frac{\epsilon}{c}<\sqrt{\beta\,\mathcal{N}\,\epsilon}\,\leq\,b_0\,\sqrt{\mathcal{N}\,\epsilon},
\] 
where $b_0=(1-\tau\,c_1^2)^{-1/2}.$  So if 
\[
\sqrt{\epsilon}<\frac{a\,\alpha}{2\,\tilde{V}\,b_0\,\sqrt{\mathcal{N}}},
\]
then we get the contradiction. And this proves the statement. 
\eprf

Thus, the point  $(c,\,\epsilon)\,\in\,\Omega$ belonging to $\partial\,\Lambda$  must be an element of the set $\Omega_1.$ However, the trajectory $q^u_{c,\,\epsilon},$ corresponding to these values ​​of the parameters, after it enters the region of small values ​​of $v$, should tend to the stationary point. This becomes obvious if we consider the following function:
\begin{equation}\label{funG_2}
\tilde{G}(v,\,u,\,w)=u^2+2\,\beta\int^v_0{f(z)d\,z}+\frac{\beta\,c}{2}\,E\,v^2-\frac{\beta}{2\,\delta}\left(2\,\delta\,\gamma+E   \right)\,w^2-2\,\beta\,v\,w+E\,u\,v.
\end{equation}
The function (\ref{funG_2}) is monotone  if $0<v<v_{min}$ and $0<E<-2\,\delta\,v_{min}/f(v_{min}).$ Therefore the orbit $q^u_{c,\,\epsilon}(\xi)\,\in\,\Omega_1 \cap \partial \Lambda$ should tend to the origin as $\xi$ tends to $+\infty.$

\begin{figure}
\begin{center}
\includegraphics[totalheight=1.9 in]{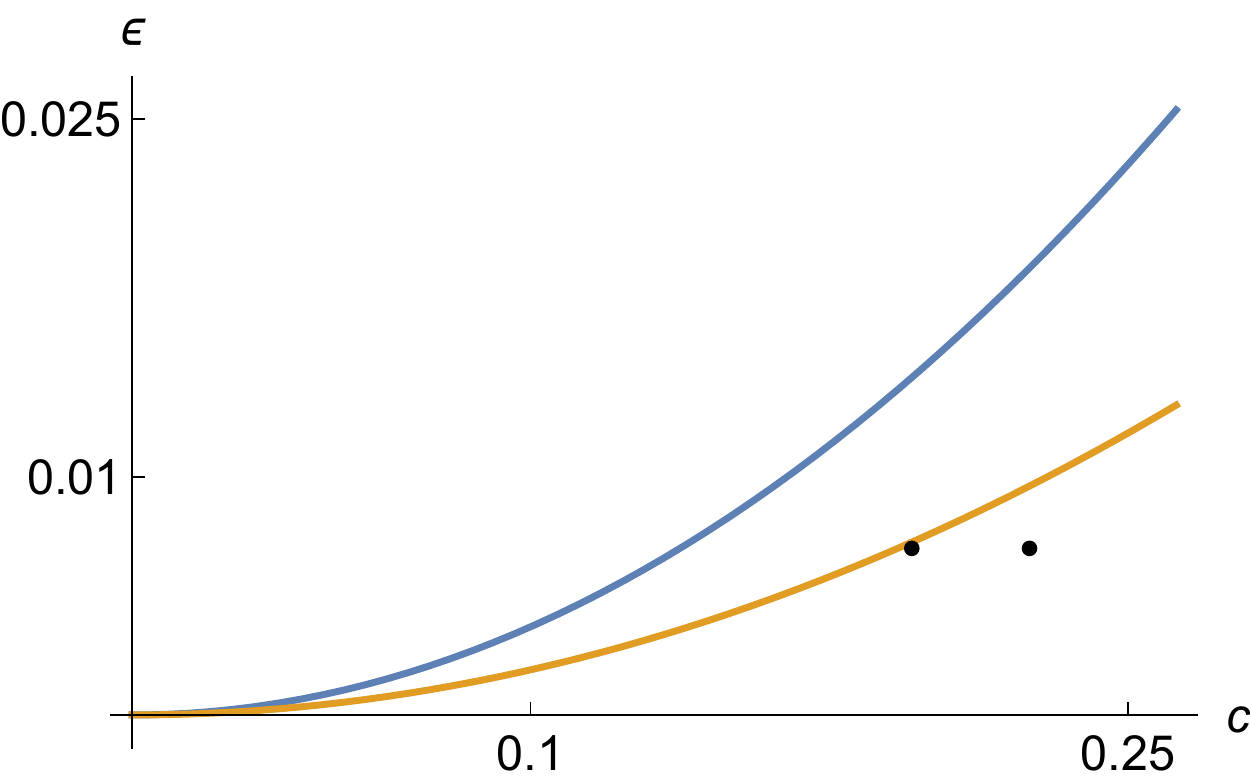}\hspace{0.5 cm}
\includegraphics[totalheight=1.9 in]{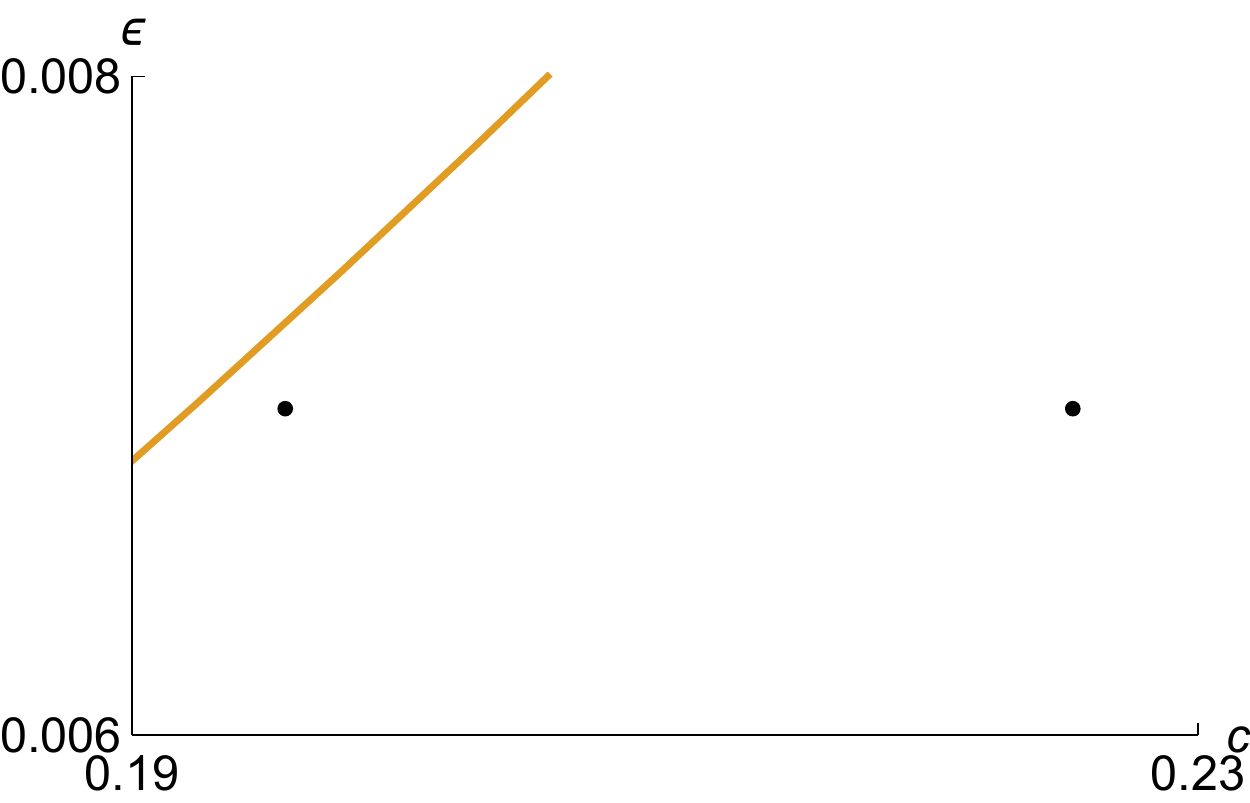}
\caption{The parameters' values corresponding to the homoclinic loops as well as the boundaries of the set $\Omega_2$ (left panel) and the enlargement of domain which contains the points (filled circles) corresponding to the solitary waves (right panel)}\label{figure5AB}
\end{center}
\end{figure}%Figure5
\begin{figure}
\begin{center}
\includegraphics[totalheight=1.5 in]{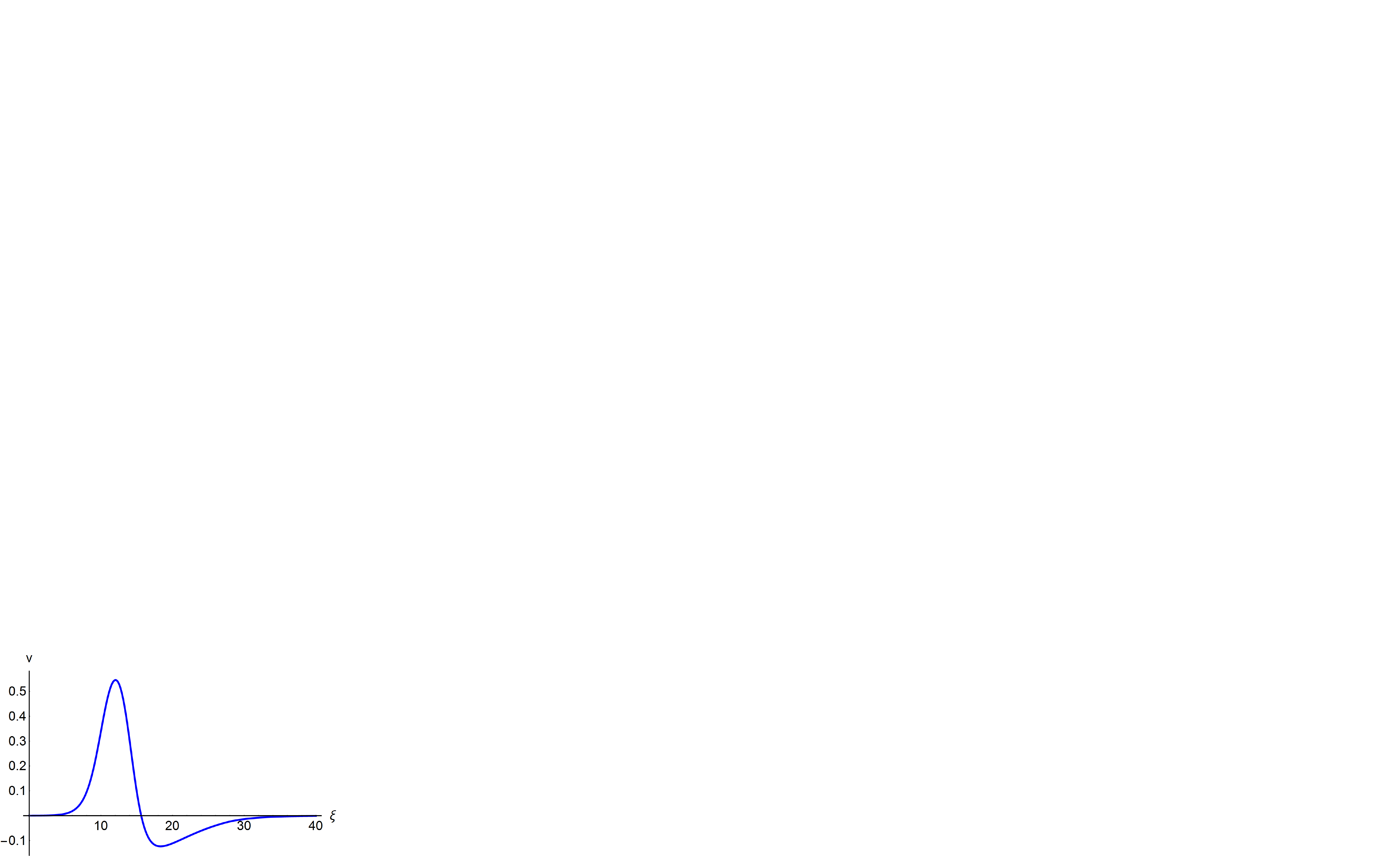}\hspace{0.5 cm}
\includegraphics[totalheight=1.5 in]{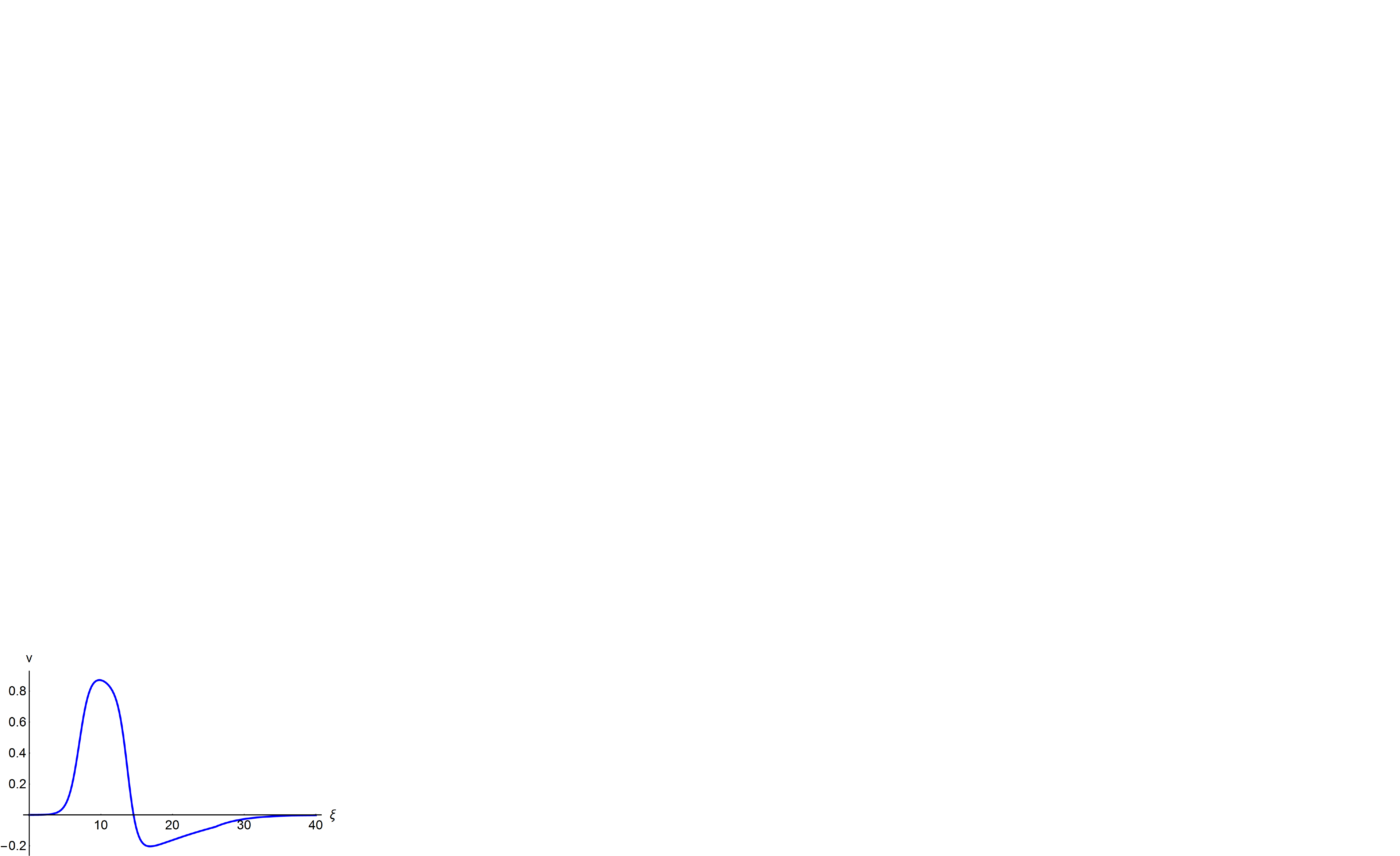}\\
(a) \hspace{4 cm} (b)
\caption{The solutions of system (\ref{uklad2}) at $c=0.195747986$ corresponding to the slow solitary wave (a) and $c=0.225305407$ relating to the fast solitary wave (b). }\label{figure6}
\end{center}
\end{figure}%Figure6

\begin{figure}
\begin{center}
\includegraphics[totalheight=1.5 in]{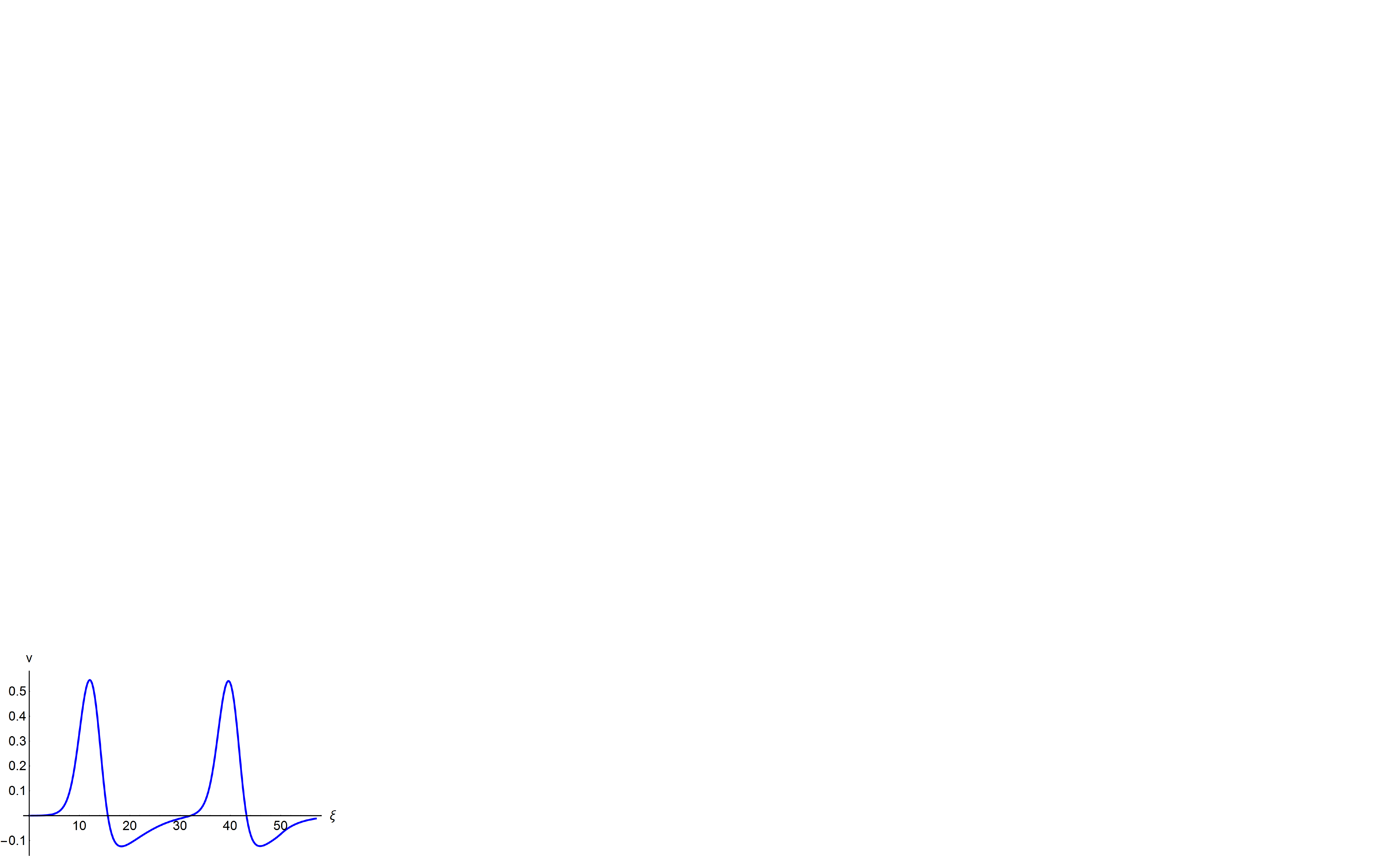}
\caption{Two-humped solitary wave solution of the system (\ref{uklad2}) obtained at  $c=0.195747986011335$ }\label{figure7}
\end{center}
\end{figure}%Figure7

\begin{rem}
The presence of homoclinic trajectories among the solutions of system (\ref{uklad2}) can be shown using the Melnikov method \cite{GH,Melnikov} if the relations $0<\epsilon \ll c \ll 1$ hold. Indeed, in this case, the system (\ref{uklad2}), up to $O(c^2)$, can be presented in the form of a perturbed Hamiltonian system:
\begin{equation}\label{ham_pert2}
\left(\begin{array}{c} v \\ u \end{array} \right)^\prime =F+K,
\end{equation} 
where $F=(u,\,-f(v))^{tr},$ $K=(0,\,c\,u+\delta\,w_1)^{tr},$ and $w_1$, up to $O(\delta^2)$, is given by the formula (\ref{W_approx}). 
So the Melnikov integral will take the form
\[ 
M=\int_{-\infty}^{+\infty} {F^{\bot}\,K d\,\xi}=c\,\int_{-\infty}^{+\infty} {u^2(\xi) d\,\xi}-\delta\,\int_{-\infty}^{+\infty} {v^2(\xi)\, d\,\xi}.
\]
It is quite obvious that at a certain ratio between $c$ and $\epsilon$ the right-hand side will be equal to zero. There is nothing surprising that the Melnikov method catches only one homoclinic trajectory for a fixed value of $\epsilon$. This is due to the fact that for the  value of the parameter $c$  lying  close to the line $c = c_1$, the assertion $0<c\,\ll\,1$ may be false.
\end{rem}

The conclusions of theoretical studies indicating the presence of a pair of homoclinic trajectories under the above restrictions on the parameters are verified using numerical simulation. Numerical experiments conducted at
$\epsilon = 0.006991097526935545$, $\gamma=2.706215020212898$,
$a=0.13$, and $\tau= 14.554975027077534$ confirm the existence of a pair of homoclinic solutions corresponding to $c=0.195747986$ (slow solitary wave) and $c=0.225305407$ (fast solitary wave). The corresponding points are located in the plane $\Omega$ to the right of the line $\epsilon=\mathcal{N}\,c^2$ (see Fig.~\ref{figure5AB}), which fully agrees with the theoretical results. Figure \ref{figure6} shows the graphs of slow and fast solitons on the physical plane. Figure \ref{figure7} demonstrates two-humped solitary wave obtained for the value $c=0.195747986011335$ located in vicinity of the value $c=0.195747986$ corresponding to the slow solitary wave.

\section{Conclusion}

 It was proved in this paper, that the system (\ref{PDEq}), under certain restrictions on the parameters, possesses a pair of soliton-like traveling wave solutions. These solutions correspond to the homoclinic trajectories  of the factorized system (\ref{uklad2}).
The existence of the solitary wave solutions is confirmed by numerical experiments, the results of which completely agree with the conclusions of the  theoretical analysis of the system (\ref{uklad2}). In numerical experiments, in addition to the presence of a simple solitary wave solutions predicted theoretically, a two-humped wave structure is found. The presence of such structures is discussed in a number of works \cite{Feroe_83,Feroe_93,GonTurGas}. They necessarily appear in cases where the main homoclinic trajectory (which can be conventionally called a one-humped structure) is doubly-asymptotic to a saddle-focus in which the condition, formulated for the first time by L.P. Shilnikov, is fulfilled \cite{Silnikov,Gaspard_93,GonTurGas,GH}. Our further efforts will be purposed at studying the stability of the solitary wave solutions found, their dynamic properties, as well as identifying the presence of multi-hump wave patterns and investigating their properties.
 
 \section*{ Acknowledgements.} The investigations carried out by two authors (A.G. and V.V.)  were partially supported by the Faculty of Applied Mathematics AGH UST within subsidy of Ministry of Science and Higher Education of Poland. S.S. greatly acknowledges warm hospitality extended to him in the course of his visit to AGH UST in Krakow.

\end{document}